\documentclass[12pt]{article}
\usepackage{amssymb, latexsym, amsmath, hyperref}
\usepackage{epsfig}
\usepackage{graphicx}
\usepackage{amsfonts}
\usepackage{mathrsfs}
\usepackage{multirow}
\usepackage[dvips]{color}

\newcommand{\C}{\mathbb{C}}

\newcommand{\fn}{{\,\mathfrak{n}\,}}

\newcommand{\fz}{\mathfrak{z}}

\newcommand{\cA}{{\mathcal{A}}}

\newcommand{\be}{\begin{equation}}
\newcommand{\ee}{\end{equation}}
\newcommand{\bea}{\begin{eqnarray}}
\newcommand{\eea}{\end{eqnarray}}
\newcommand{\nn}{\nonumber}

\newcommand{\ed}{\end{document}}

\newcommand{\rz}{{\mbox{\scriptsize${\rm Z}$}}}

\newcommand{\bi}{\begin{itemize}}
\newcommand{\ei}{\end{itemize}}

\newcommand{\bce}{\begin{center}}
\newcommand{\ece}{\end{center}}

\newcommand{\RE}{{\rm Re}}
\newcommand{\IM}{{\rm Im}}

\begin{document}

\title{Spectral Singularities of a Complex Spherical Barrier Potential and
Their Optical Realization}

\author{Ali~Mostafazadeh\thanks{E-mail address:
amostafazadeh@ku.edu.tr, Phone: +90 212 338 1462, Fax: +90 212 338
1559}~~and Mustafa Sar{\i}saman\thanks{E-mail address:
msarisaman@ku.edu.tr, Phone: +90 212 338 1378, Fax: +90 212 338
1559}
\\
Department of Mathematics, Ko\c{c} University, \\ 34450 Sar{\i}yer,
Istanbul, Turkey}

\date{ }
\maketitle

\begin{abstract}
The mathematical notion of a spectral singularity admits a physical
interpretation as a zero-width resonance. It finds an optical realization as a certain type of lasing effect that occurs at the threshold gain. We explore spectral singularities of a complex spherical barrier potential and study their realization as transverse spherical electromagnetic waves emitted by a gain medium with a spherical geometry. In particular, for a typical dye laser material, we obtain a lower bound on the size of the gain medium for the occurence of this kind of spectral singularities.
\vspace{2mm}

\vspace{2mm}

\end{abstract}

An interesting by-product of the recent study of complex potentials
supporting a real spectrum \cite{bender-review, review} is the
discovery of the optical realizations
\cite{prl-2009,pra-2009,pra-2011,p98} of the mathematical notion of
a spectral singularity \cite{ss}. For a complex scattering potential
defined on the real line, these are certain points of the
continuous spectrum of the corresponding non-Hermitian Schr\"odinger
operator at which the reflection and transmission
coefficients diverge \cite{prl-2009}. As a result, they correspond
to scattering states that behave exactly like resonances. This
observation has also found applications in condensed matter systems
\cite{longhi} and triggered further study of the subject
\cite{others}.

The study of the optical realizations and applications of spectral
singularities has so far been confined to effectively
one-dimensional models involving infinitely long waveguides
\cite{prl-2009,pra-2009} or infinite planar slab gain media
\cite{pra-2011,p98}. In the present article we examine spectral
singularities of a three-dimensional complex spherical barrier
potential that admits a physical realization involving an optical
gain medium with a spherical geometry.

Consider the propagation of electromagnetic (EM) fields in a linear
source-free  isotropic dielectric with time-independent (complex) permittivity $\varepsilon= \varepsilon(\vec{r})$ and constant permeability $\mu = \mu_{0}$. The
electric and magnetic fields, $\vec{E}$ and $\vec{B}$, satisfy
Maxwell's equations
    \begin{equation}
    \vec{\nabla}\cdot\vec{D} = 0, \ \ \ \ \ \ \ \ \ \ \ \ \ \ \ \ \
    \vec{\nabla}\cdot\vec{B} = 0, \label{equation1}
    \end{equation}
    \begin{equation}
    \vec{\nabla} \times \vec{E} + \dot{\vec{B}} = 0, \ \ \ \ \ \ \ \ \ \
    \ \vec{\nabla} \times \vec{H} - \dot{\vec{D}} = 0,
    \label{equation2}
    \end{equation}
where $\vec{D} := \varepsilon(\vec{r}) \vec{E}$, $\vec{H} := \mu_0
\vec{B}$, and a dot represents the time-derivative. Following the
standard derivation of the wave equation, we can use
(\ref{equation2}) to obtain
    \begin{equation}
    \ddot{\vec{E}} + \Omega^{2} \vec{E} = 0
    \label{equation3}
    \end{equation}
where $\Omega^{2} :=[\mu_{0}\varepsilon(\vec{r})]^{-1}\vec{\nabla}
\times \vec{\nabla}$.

For a time-harmonic EM field propagating in a spherically symmetric
dielectric we have $\vec{E}(\vec{r}, t) = e^{-i\omega t}
\vec{E}(\vec{r}, t)$ and $\varepsilon(\vec{r}) = \varepsilon_{0}
\mathfrak{z} (r)$, where $r$ is the radial spherical coordinate and \linebreak $\fz:[0,\infty)\to\C$ is a complex-valued function. In
this case (\ref{equation3}) reduces to the time-independent
Schr\"odinger equation:
    \begin{equation}
    -\nabla^{2} \vec{E}(\vec{r}) +  \vartheta (r) \vec{E}(\vec{r})
    = k^{2} \vec{E}(\vec{r}),
    \label{equation4}
    \end{equation}
where  $k := w/c$ and $\vartheta (r):=k^{2}[1-\mathfrak{z}(r)]$.
The latter is a spherical complex barrier potential provided that the
dielectric medium is confined in a spherical region. This is the
case for
    \be
    \fz(r)=\left\{\begin{array}{ccc}
    \fn^2 & {\rm for} & r<a,\\
    1 & {\rm for} & r\geq a,
    \end{array}\right.~~~~~~~
    \vartheta(r)=\left\{\begin{array}{ccc}
    k^2(1-\fn^2 ) & {\rm for} & r<a,\\
    0 & {\rm for} & r\geq a,
    \end{array}\right.
    \ee
where $\fn$ is the complex refractive index of the dielectric medium that is supposed to be constant.

Consider spherically symmetric solutions of (\ref{equation4}) that
define transverse radially propagating spherical waves. They have
the general form $\vec{E} (\vec{r}) = E_{\theta}(r) \hat{\theta} +
E_{\phi}(r) \hat{\phi}$, where $\hat\theta$ and $\hat\phi$ are the unit vectors along the $\theta$- and $\phi$-directions. Inserting this ansatz in (\ref{equation4}) we find that either $\theta=\pi/2$ or $E_{\theta} = 0$. In what
follows we consider the latter case, so that $\vec{E} (\vec{r}) =
E_{\phi}(r) \hat{\phi}$, and (\ref{equation4}) reduces to
    \begin{equation}
    \left[\frac{d^{2}}{dr^{2}} + \frac{2}{r}\frac{d}{dr} + k^{2}
    -\vartheta(r) -\frac{1}{r^{2}}\right]E_{\phi} = 0.
    \label{equation5}
    \end{equation}
For $r<a$ and $r\geq a$ where $\vartheta$ takes constant values, we can transform (\ref{equation5}) to the Bessel equation of order $\nu:= \sqrt{5}/2$, \cite{jackson}. The electric field inside and outside the dielectric region are respectively given by
    \begin{equation}
    \vec{E}_{\rm outside} = [A_{1} h_\nu^{(1)} (kr) + A_{2} h_\nu^{(2)}
    (kr)]\hat{\phi},
    \label{equation6}
    \end{equation}
    \begin{equation}
    \vec{E}_{\rm inside} = [B_{1} j_\nu(\tilde{k}r) + B_{2} n_\nu
    (\tilde{k}r)]\hat{\phi},
    \label{equation7}
    \end{equation}
where the coefficients $A_{1, 2}$ and $B_{1, 2}$ are related via the
appropriate boundary conditions at $r=a$, $j_\nu$ and
$h_\nu^{(1, 2)}$ are respectively the spherical Bessel and Hankel
functions, and $\tilde{k} := \fn\,k$. Demanding the
electric field to disappear at the origin, we have $B_{2}=0$ and
    \begin{equation}
    \vec{E}_{inside} = B_{1} j_\nu(\tilde{k}r)\,\hat{\phi}.
    \label{equation8}
    \end{equation}

In view of the asymptotic expression for the Hankel functions, namely
    \[ h_\nu^{(1)}(\rz)=\frac{e^{i\rz}}{\rz}\left[-ie^{\frac{-i\pi \nu}{2}}+{\cal O}({\rz}^{-1})\right],~~~~~
    h_\nu^{(2)}(\rz)=\frac{e^{-i\rz}}{\rz}\left[ie^{\frac{i\pi \nu}{2}}+{\cal O}({\rz}^{-1})\right],\]
the terms on the right-hand side of (\ref{equation6}) correspond to the reflected and incident spherical waves, respectively. As a result, we identify $R:=|A_{1}/A_{2}|^2$ with the reflection coefficient of the system whose real and positive poles in the $k^2$-plane are the spectral singularities. In order to determine these poles, we need to derive the relationship between the coefficients $A_{1, 2}$ and $B_{1}$. First, we use (\ref{equation2}) to compute the magnetic field:
    \begin{equation}
    \vec{B}_{\rm outside} = \frac{i}{\omega}
    \left[A_1\left\{\frac{h_\nu^{(1)} (kr)}{r}+\frac{dh_\nu^{(1)}(kr)}{dr}\right\}+ A_{2}    \left\{\frac{h_\nu^{(2)}(kr)}{r} +\frac{dh_\nu^{(2)}(kr)}{dr}\right\}\right] \hat{\theta},
    \label{equation9}
    \end{equation}
    \begin{equation}
    \vec{B}_{\rm inside} = \frac{iB_1}{\omega}\left[\frac{j_\nu (\tilde{k}r)}{r} +    \frac{dj_\nu(\tilde{k}r)}{dr}\right]  \hat{\theta}.
    \label{equation10}
    \end{equation}
Imposing the standard boundary conditions:
$(\vec{E}_{\parallel})_{\rm inside} = (\vec{E}_{\parallel})_{\rm outside}$,
$(\vec{H}_{\parallel})_{\rm inside} =
(\vec{H}_{\parallel})_{\rm outside}$ and noting that $\vec H=\mu_0\vec B$, we have
    \bea
    &&B_{1}j_\nu(\tilde{k}a) = A_{1} h_\nu^{(1)}(ka) + A_{2} h_\nu^{(2)} (ka)
    \label{equation11},\\
    &&B_{1}\left[\frac{j_\nu(\tilde{k}a)}{a} +
    \frac{dj_\nu(\tilde{k}r)}{dr}\Big|_{r = a}\right] =
     A_1\left[\frac{h_\nu^{(1)}(k a)}{a} +\frac{dh_\nu^{(1)}(k r)}{dr}\Big|_{r=a}\right]+
    A_{2}\left[\frac{h_\nu^{(2)}(k a)}{a} +
    \frac{dh_\nu^{(2)}(k r)}{dr}\Big|_{r=a}\right].~~~~~
    \label{equation12}
    \eea
We can easily eliminate $B_{1}$ in (\ref{equation11}) and (\ref{equation12}), and compute the reflection amplitude:
    \be
    \frac{A_1}{A_2}=\frac{h_{\nu}^{(2)}(ka)\left[\frac{j_\nu (\tilde{k}a)}{a} +
    \frac{dj_\nu (\tilde{k}r)}{dr}\Big|_{r = a}\right]-
    j_\nu (\tilde{k}a) \left[\frac{h_{\nu}^{(2)}(ka)}{a} +
    \frac{dh_{\nu}^{(2)}(kr)}{dr}\Big|_{r=a}\right]}{
    j_\nu (\tilde{k}a)\left[\frac{h_{\nu}^{(1)}(ka)}{a} +\frac{dh_{\nu}^{(1)}(kr)}{dr}\Big|_{r=a}\right]-
     h_{\nu}^{(1)}(ka)\left[\frac{j_\nu (\tilde{k}a)}{a} +
    \frac{dj_\nu (\tilde{k}r)}{dr}\Big|_{r = a}\right]}
     \ee
Setting the denominator of this quantity equal to zero we find the following condition for the existence of a spectral singularity.
    \begin{equation}
    \frac{d}{dr} \ln h_{\nu}^{(1)}(kr)\Big|_{r=a} = \frac{d}{dr} \ln
    j_\nu (\tilde{k}r)\Big|_{r=a}.
    \label{equation14}
    \end{equation}

Recalling the recursion formula \cite{jackson} for $h_{\nu}^{(1)}$ and $j_\nu $:
    \begin{equation}
    \frac{du_\nu(\rz)}{d\rz}= \frac{\nu \,u_{\nu-1}(\rz) - (\nu+1)u_{\nu+1}(\rz)}{2\nu + 1},
    \label{recur}
    \end{equation}
with $u_\nu=j_\nu $ or $h_{\nu}^{(1)}$, we can express (\ref{equation14}) in the form
    \begin{equation}
    \frac{[\nu h_{\nu-1}^{(1)}(ka)-(\nu+1)h_{\nu+1}^{(1)} (ka)]}{h_{\nu}^{(1)}(ka)}
    = \frac{\fn[\nu j_{\nu-1}(\tilde{k}a)-(\nu+1)j_{\nu+1}(\tilde{k}a)]}{j_\nu (\tilde{k}a)},
    \label{equation15}
    \end{equation}
where we have used the fact that $\tilde k=\fn k$.
Because $\fn$ and consequently $\tilde k$ take complex values, this is a
complex equation. Equating the real and imaginary parts of both sides of this equation we find two real equations for the three real variables $\eta:=\RE(\fn)$, $\kappa:=\IM(\fn)$, and $x:=ka$. Because the determination of the explicit form of these equations is not easy, we examine their asymptotic expressions that are valid for $x\gg 1$. For typical situations the radius of the dielectric ball is much larger than the wavelength of the wave. Therefore, for practical purposes that we will consider, the condition $x\gg 1$ holds and the asymptotic treatment of (\ref{equation15}) provides extremely reliable results.

The asymptotic expansion of the spherical Bessel and Hankel functions, $j_\nu$ and $h^{(1)}_\nu$, that are valid in the large argument limit ($|\rz|\gg 1$), have the form
     \bea
     j_\nu (\rz) &=& \frac{\sin(\rz - \frac{\pi\nu}{2})}{\rz}
     \sum_{s=0}^{\infty} \frac{(-1)^{s}\cA_{2s}(\nu)}{\rz^{2s}} + \frac{\cos(z-\frac{\pi\nu}{2})}{\rz} \sum_{s=0}^{\infty} \frac{(-1)^{s}\cA_{2s+1}(\nu)}{\rz^{2s+1}},
     \label{asymp-j}\\
     h_{\nu}^{(1)}(\rz)&=& \frac{e^{i(\rz-\frac{\pi\nu}{2})}}{\rz}  \left[ - i \sum_{s=0}^{\infty} \frac{(-)^{s}\cA_{2s}(\nu)}{\rz^{2s}}+
     \sum_{s=0}^{\infty} \frac{(-1)^{s}\cA_{2s+1}(\nu)}{\rz^{2s+1}}\right], \label{asymp-h}
     \eea
where
    \be
    \cA_k(\nu):=\frac{\Gamma(\nu+k+1)}{2^{k} k! \Gamma(\nu-k+1)}=
    \frac{\prod_{\ell=0}^{2k-1}(\nu+k-\ell)}{2^{k} k!},\nn
    \ee
and $\Gamma$ stands for the Gamma function. Substituting (\ref{asymp-j}) and (\ref{asymp-h}) in (\ref{equation15}), neglecting the quadratic and higher order terms in $x^{-1}$ and noting that $\cA_0(\nu)=1$ and $\cA_1:=\cA_1(\nu)=\frac{\nu(\nu+1)}{2}$, we find
    \be
    \tan(\fn x-\frac{\pi\nu}{2})\approx-i\fn+\frac{(\fn^2-1)\cA_1}{\fn\, x}.
    \label{ss-eq-1}
    \ee

Next, we wish to compute the real and imaginary parts of (\ref{ss-eq-1}) and express them in terms of the real parameters $\eta,\kappa$ and $x$. To this end, first we recall the identity:
    \be
    \tan^{-1}(\rz)=\pi m+\frac{1}{2i}\ln\left(\frac{1+i\rz}{1-i\rz}\right),
    \label{identity}
    \ee
where $\rz$ is a complex variable, $m$ is an integer, and ``$\ln$'' denotes the principal part of the natural logarithm. If we set $\rz=-i\fn+\frac{A_1(\fn^2-1)}{\fn\, x}$ and use (\ref{ss-eq-1}) and (\ref{identity}), we obtain
    \be
    \fn x-\frac{\pi\nu}{2}\approx \pi m+\frac{1}{2i}\left[\ln\left(\frac{\fn+1}{\fn-1}\right)+
    \ln\left(-1+\frac{2i\cA_1}{\fn x}\right)\right].
    \label{eq-ss-2}
    \ee
Note that
    \[\ln\left(-1+\frac{2i\cA_1}{\fn x}\right)=\ln(-1)+\ln\left(1-\frac{2i\cA_1}{\fn x}\right)\approx \pi i-\frac{2i\cA_1}{\fn x}.\]
Therefore, if we only keep the first two largest powers of $x$ in (\ref{eq-ss-2}), the term proportional to $\cA_1$ drops out of the calculations and we obtain
    \be
    \fn x\approx \pi\left(m+\frac{\nu+1}{2}\right)+
    \frac{1}{2i}\ln\left(\frac{\fn+1}{\fn-1}\right).
    \label{eq-ss-3}
    \ee

Next, we employ $\fn=\eta+i\kappa$ to obtain
    \be
    \frac{\fn+1}{\fn-1}=\sqrt{\frac{(\eta+1)^2+\kappa^2}{(\eta-1)^2+\kappa^2}}
    \,\exp\left[i\arctan\left(\frac{-2\kappa}{\eta^2+\kappa^2-1}\right)\right]\approx
   \left(\frac{\eta+1}{\eta-1}\right)
   \exp\left(\frac{-i2\kappa}{\eta^2-1}\right),
   \label{id3}
    \ee
where ``$\arctan$'' stands for the principal part of ``$\tan^{-1}$'' with values in $(-\frac{\pi}{2},\frac{\pi}{2})$. Furthermore, because for realistic gain media $\kappa\ll 1\approx\eta$, we neglect the quadratic and higher order terms in $\kappa$. Substituting $\fn=\eta+i\kappa$ and (\ref{id3}) in (\ref{eq-ss-3}), we arrive at
    \bea
    x\,\eta &\approx& \pi\left(m+\frac{\nu+1}{2}\right),
    \label{ss-5a}\\
    x\,\kappa&\approx&-\frac{1}{2}\,\ln\left(\frac{\eta+1}{\eta-1}\right).
    \label{ss-5b}
    \eea
Equivalently,
    \bea
    \kappa& \approx &\frac{-\eta\ln\left(\frac{\eta+1}{\eta-1}
    \right)}{\pi\left(2m+\nu+1\right)},
    \label{ss-6a}\\
    x& \approx &-\frac{1}{2\kappa}\,\ln\left(\frac{\eta+1}{\eta-1}\right).
    \label{ss-6b}
    \eea
Because $\eta$ and $x$ take positive values and $x\gg 1$, the integer $m$ that we view as a mode number take large positive integer values. This implies that $\kappa<0$. Therefore, in order to create a spectral singularity the dielectric must consist of a gain medium. This is an intriguing mathematical manifestation of the law of conservation of energy. The same results is obtained in \cite{pra-2009,pra-2011} while studying the optical spectral singularities of a one-dimensional waveguide including a gain region and an infinite planar slab gain medium.

Equation~(\ref{ss-6a}) determines the approximate location of the spectral singularities in the $\eta$-$\kappa$ plane (complex $\fn$-plane). It yields an infinite family of curves parameterized by the mode number $m$. Given that $x=ak$, for each point $(\eta_\star,\kappa_\star)$ that lies on one of the spectral singularity curves, Equation~(\ref{ss-6b}) specifies the value of $k^2$ for the corresponding spectral singularities.

In practice the refractive index $\fn$ depends on the properties of the gain medium and the wavelength $\lambda:=2\pi/k=2\pi c/\omega$  of the propagating EM wave. In order to realize an optical spectral singularity, we need to impose the dispersion relation that determines the dependence of $\fn$ on $\lambda$ and other relevant physical parameters.

Following \cite{pra-2011}, we consider a gain medium that is obtained by doping a host medium of refraction index $n_0$ and modeled by a two-level atomic system with lower and upper level population densities $N_l$ and $N_u$, resonance frequency $\omega_0$, and damping coefficient $\gamma$. Then the dispersion relation takes the form
    \be
    \fn^2= n_0^2-
    \frac{\hat\omega_p^2}{\hat\omega^2-1+i\hat\gamma\,\hat\omega},
    \label{epsilon}
    \ee
where $\hat\omega:=\omega/\omega_0$, $\hat\gamma:=\gamma/\omega_0$,
$\hat\omega_p:=(N_l-N_u)e^2/(m_e\varepsilon_0\omega_0^2)$, $e$ is electron's charge, $m_e$ is its mass, and $\varepsilon_0$ is the permittivity of the vacuum.
We can express $\hat\omega_p^2$ in terms of the imaginary part $\kappa_0$ of $\fn$ at the resonance wavelength $\lambda_0:=2\pi c/\omega_0$ according to \cite{pra-2011}
    \be
    \hat\omega_p^2\approx2n_0\hat\gamma\kappa_0,
    \label{plasma}
    \ee
where the approximation symbol means that we neglect quadratic terms in $\kappa_0$.

Inserting~(\ref{plasma}) in (\ref{epsilon}) and using $\fn=\eta+i\kappa$, we obtain
    \be
    \eta\approx n_0+\kappa_0f_1(\hat\gamma,\hat\omega) ,~~~~
    \kappa\approx\kappa_0f_2(\hat\gamma,\hat\omega) ,
    \label{eqz301}
    \ee
where
    \be
    f_1(\hat\gamma,\hat\omega):=\frac{\hat\gamma(1-\hat\omega^2)}{(1-\hat\omega^2)^2+
    \hat\gamma^2\hat\omega^2},~~~~
    f_2(\hat\gamma,\hat\omega):=\frac{\hat\gamma^2\hat\omega}{(1-\hat\omega^2)^2+
    \hat\gamma^2\hat\omega^2}.
    \label{fs}
    \ee
Now, we substitute (\ref{eqz301}) in (\ref{ss-5a}), recall that $x=ka=2\pi a/\lambda$, and keep the leading order term in the small $\kappa_0$ approximation. This yields
    \be
    \lambda\approx \frac{4n_0a}{2m+\nu+1}.
    \label{lambda}
    \ee
Next, we recall that $\kappa_0$ may be related to the gain coefficient $g_0$ at the resonance frequency according to $g_0=-4\pi\kappa_0/\lambda_0$. Using this relation and Equations~(\ref{ss-5a}) and (\ref{eqz301}), we obtain
    \be
    g_0\approx\frac{4n_0\ln\left(\frac{n_0+1}{n_0-1}\right)}{
    \lambda_0(2m+\nu+1)f_2(\hat\gamma,\lambda_0/\lambda)}.
    \label{gain}
    \ee
In view of (\ref{lambda}) this relation takes the form
    \be
     g_0\approx\frac{1}{a}\ln\left(\frac{n_0+1}{n_0-1}\right)
     \left[1+\frac{f(\lambda)}{\hat\gamma^2}\right],~~~~~~~
    f(\lambda):=\left(\frac{\lambda^2-\lambda_0^2}{\lambda_0\lambda}\right)^2.
    \label{gain2}
    \ee
These equations describe the dependence of the gain coefficient $g_0$ necessary for generating a spectral singularity on the radius $a$ of the spherical gain medium, the normalized damping coefficient $\hat\gamma$, and the wavelength $\lambda$. $g_0$ is inversely proportional to $a$. Its dependence on $\lambda$ is given by the function $f$ which has a minimum at the resonance wavelength $\lambda_0$ and increases monotonically as $|\lambda-\lambda_0|$ increases. This shows that the spectral singularity with the wavelength closest to the resonance wavelength of the gain medium requires the smallest necessary gain. This is in complete agreement with the results of \cite{p98}.

Because our model involves no mirrors, we need a high-gain medium to create a spectral singularity. In the following we consider a semiconductor diode laser (disregarding the difficulty of manufacturing a spherical diode) and a dye laser.

Consider a spherical diode laser with the following characteristics \cite{silfvast}:
    \be
    n_0=3.4,~~~\lambda_0=1500\,{\rm nm},~~~
    \hat\gamma=0.02,~~~g_0\leq 1000\,{\rm cm}^{-1}.
    \label{specific}
    \ee
Then according to (\ref{lambda}) the spectral singularity with wavelength closest to the resonance wavelength $\lambda_0$ corresponds to the mode number $m=679$, if we take $a=150~\mu{\rm m}$. Its wavelength and the required gain coefficient are respectively given by $\lambda=1499.870~{\rm  nm}$ and $g_0=40.412~{\rm cm}^{-1}$. There are a total of 66 spectral singularities with mode numbers in the range 647-712 whose realization requires a gain coefficient less that $1000\,{\rm cm}^{-1}$. Their wavelength decreases from $1573.930~{\rm  nm}$ to $1430.457~{\rm  nm}$ as $m$ takes values from $647$ to $712$.

\begin{figure}
    \begin{center}
    \includegraphics[scale=1,clip]{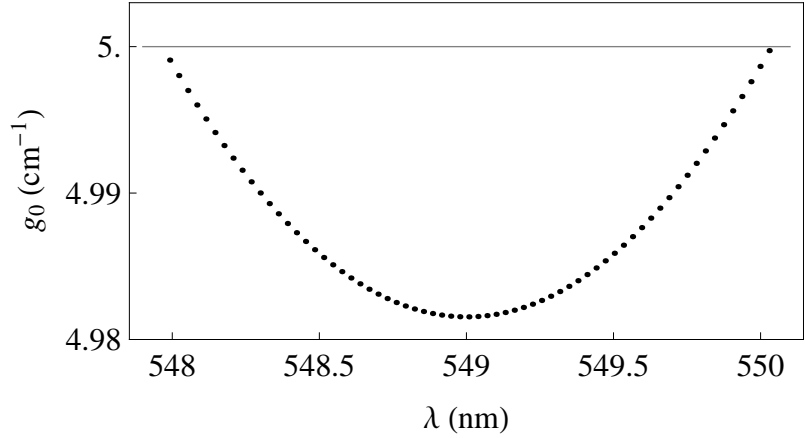}
    \caption{Spectral singularities of a spherical dye gain medium with specifics~(\ref{specific-dye}) and radius $a=3.300~{\rm mm}$. The mode number $m$ ranges over $17746$-$17812$ as the wavelength $\lambda$ decreases. The spectral singularity requiring the least gain coefficient $g_0$ has a wavelength that is closest to the resonance wavelength of the medium. This corresponds to $m=17779$. The grey horizontal line represents the experimental upper limit on $g_0$.\label{fig1}}
    \end{center}
    \end{figure}

A more realistic choice for the gain medium that supports the above-examined spectral singularities is a spherical dye laser. For definiteness consider a Rose Bengal-DMSO (Dimethyl sulfoxide) solution with the following characteristics \cite{silfvast,dye}.
    \be
    n_0=1.479,~~~\lambda_0=549\,{\rm nm},~~~
    \hat\gamma=0.062,~~~g_0\leq 5\,{\rm cm}^{-1}.
    \label{specific-dye}
    \ee
Using these figures in (\ref{lambda}) and (\ref{gain}) we find out that the smallest value of radius $a$ that supports a spectral singularity (and highest gain coefficient $g_0=5\,{\rm cm}^{-1}$) is $3.287825~{\rm mm}$ which is within the experimentally attainable range \cite{silfvast}. For a slightly larger sample with radius $a=3.300~{\rm mm}$, we can generate 67 different spectral singularities by varying the gain coefficients in the range $4.981546-4.999727\,{\rm cm}^{-1}$. These correspond to the mode numbers $17746$-$17812$. As $m$ takes values in this range, the wavelength of the corresponding spectral singularities decreases from $550.028673~{\rm  nm}$ to $547.991700~{\rm  nm}$. Figure~\ref{fig1} shows the location of these spectral singularities in the $\lambda$-$g_0$ plane.
The dependence of the quantities $m$, $\lambda$ and $g_0$ on one another is similar to that of the spectral singularities of the one-dimensional system studied in \cite{pra-2011,p98}. As one starts pumping the gain medium there arises no spectral singularities unless the gain coefficient reaches its first critical value ($g_0^{(1)}=4.981546~{\rm cm}^{-1}$.) This is the gain required to excite the spectral singularity with a wavelength  ($\lambda^{(1)}=549.008297507~{\rm nm}$) that is closest to the resonance wavelength of the medium. As $g_0$ increases further this spectral singularity disappears and there are no other spectral singularities until $g_0$ takes its second critical value, namely $g_0^{(2)}=4.981554~{\rm cm}^{-1}$. This excites the spectral singularity with wavelength $\lambda^{(2)}=548.977436136~{\rm nm}$. This process continues until one saturates the upper experimental bound on the gain coefficient. As noted in \cite{pra-2011,p98}, it can be used to generate a tunable laser whose wavelength may be adjusted by changing the pumping intensity.

Table~\ref{table1} gives the values of the first seven critical values of $g_0$ and $\lambda$ that we have obtained using perturbative and exact (numerical) calculations. The extremely good agreement between the perturbative and exact results is related to the fact the perturbative calculations amount to ignoring terms of order $x^{-2}$ and $\kappa_0^2$. For the dye laser we consider these are about $7\times 10^{-10}$ and $5\times 10^{-10}$, respectively.
                \begin{table}
                \begin{center}
                \begin{tabular}{|c|c|c|c|c|}
                \hline
                $\ell$ & $g_0^{(\ell)}({\rm cm}^{-1})$ & $m$ & 
                $\lambda_{{\rm pert.}}^{(\ell)}$({\rm nm}) &
                $\lambda_{{\rm exact}}^{(\ell)}$({\rm nm})  \\
                \hline
                1 & 4.981546 & 17779 & 549.00830142 & 549.00829751 \\
                2 & 4.981554 & 17780 & 548.97742540 & 548.97743614 \\
                3 & 4.981572 & 17778 & 549.03918091 & 549.03916235 \\
                4 & 4.981594 & 17781 & 548.94655285 & 548.94657824 \\
                5 & 4.981630 & 17777 & 549.07006387 & 549.07003065 \\
                6 & 4.981668 & 17782 & 548.91568378 & 548.91572380 \\
                7 & 4.981720 & 17776 & 549.10095031 & 549.10090243\\
                \hline
                \end{tabular}
                \end{center}
                \caption{The first seven critical values of the gain coefficient $g_0$ and the wavelength of the corresponding spectral singularities of the spherical dye gain medium~(\ref{specific-dye}) with radius $a=3.300~{\rm mm}$. $\lambda_{{\rm pert.}}^{(\ell)}$ and $\lambda_{{\rm exact}}^{(\ell)}$ are respectively the values of the wavelength obtained using perturbation theory and exact numerical calculations. $m$ is the mode number. 
                \label{table1}}
                \end{table}

\begin{figure}
    \begin{center}
    \includegraphics[scale=.70,clip]{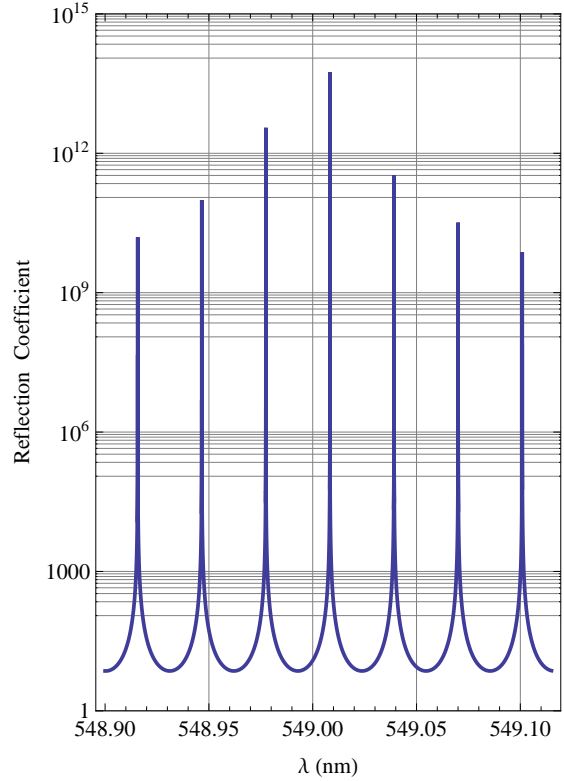}
    \caption{(Color online) Logarithmic plot of the reflection coefficient as a function of the wavelength for the spherical dye gain medium with specifics~(\ref{specific-dye}), radius $a=3.300~{\rm mm}$, and gain coefficient $g_0=g_0^{(1)}=4.981546~{\rm cm}^{-1}$ that produces the
    spectral singularity with wavelength $\lambda=\lambda^{(1)}=549.00329751~{\rm nm}$. This corresponds to the central peak. Other peaks represents resonances that turn into spectral singularities if we increase $g_0$. These have a height that is at least 100 times smaller than that of the spectral singularity at $\lambda=\lambda^{(1)}$. The peak to the left of the central peak appears at $\lambda=\lambda^{(2)}$. It is the first resonance that becomes a spectral singularity as $g_0$ increases. This explains why it has the second largest height. \label{fig2}}
    \end{center}
    \end{figure}

Figure~\ref{fig2} shows a logarithmic plot of the exact expression for the reflection coefficient, $R:=|A_1/A_2|^2$, as a function of the wavelength $\lambda$ for the above spherical dye laser with radius $a=3.300~{\rm mm}$ and gain coefficient $g_0=g_0^{(1)}$. The highest peak with $R> 5\times 10^{14}$ demonstrates the spectral singularity with  wavelength $\lambda=\lambda^{(1)}$. The other peaks with $R<4\times 10^{12}$ are the resonances that turn into spectral singularities as one increases $g_0$. If we use more and more date points to produce this graph, the height of the central peak increases indefinitely while those of the others remain essentially unchanged. This is a clear indication that the central peak corresponds to a spectral singularity.

In conclusion, we have solved the problem of finding spectral singularities of a complex spherical barrier potential and outlined a possible physical realization of the corresponding resonance effect. This is done by examining spherical transverse electromagnetic waves emitted by a spherical gain medium. We have obtained the particular values of the physical parameters of the system for which a spectral singularity appears. In particular, we have performed a reliable perturbative analysis of the problem that reveals the emergence of a mode number and leads to an explicit relationship between the necessary gain coefficient, the radius of the sample, and the wavelength of the spectral singularity. For definiteness, we examined the numerical values of these quantities for a concrete dye laser and obtained a lower bound on the radius of the sample that would support a spectral singularity. This gives $a\approx 3.29~{\rm mm}$ that is much larger than the typical mirosphere dye lasers. The reason is that the spectral singularities we study correspond to a lasing effect that is different from the lasing due to the whispering gallery modes \cite{wgm}. This is simply because we consider spherical electromagnetic waves.

\vspace{.3cm}
\noindent {\em Acknowledgments:} This work has been
supported by  the Scientific and Technological Research Council of
Turkey (T\"UB\.{I}TAK) in the framework of the project no: 110T611
and the Turkish Academy of Sciences (T\"UBA).

\end{document}